\documentclass[conference]{IEEEtran}
\IEEEoverridecommandlockouts
\usepackage{cite}
\usepackage{amsmath,amssymb,amsfonts}
\usepackage{algorithmic}
\usepackage{graphicx}
\usepackage{textcomp}
\usepackage{xcolor}
\def\BibTeX{{\rm B\kern-.05em{\sc i\kern-.025em b}\kern-.08em
    T\kern-.1667em\lower.7ex\hbox{E}\kern-.125emX}}

\usepackage{placeins}
\usepackage{tikz}

\graphicspath{{./Figures/}}
\DeclareGraphicsExtensions{.pdf,.jpeg,.png,.jpg, .PDF, .JPEG, .PNG, .JPG}

\newcommand{\putsec}[2]{\section{#1}\label{sec:#2}}
\newcommand{\putsubsec}[2]{\subsection{#1}\label{sec:#2}}
\newcommand{\figref}[1]{Figure~\ref{#1}}

\makeatletter
\def\blfootnote{\gdef\@thefnmark{}\@footnotetext}
\makeatother

\newcommand{\name}{\texttt{Softermax}}
\usepackage{multirow}
\usepackage{wrapfig}
\usepackage{subfig}
    
\begin{document}

\title{Softermax: Hardware/Software Co-Design of an Efficient Softmax for Transformers\\
\thanks{*This work was performed while an intern at NVIDIA.}
}

\author{\IEEEauthorblockN{Jacob R. Stevens$^{1*}$, Rangharajan Venkatesan$^2$, Steve Dai$^2$, Brucek Khailany$^2$, Anand Raghunathan$^1$}
\IEEEauthorblockA{
\textit{Purdue University, West Lafayette$^1$}\\
\textit{NVIDIA$^2$}\\
\{steven69, raghunathan\}@purdue.edu\\
\{rangharajanv, sdai, bkhailany\}@nvidia.com}
}
%\and
%\IEEEauthorblockN{2\textsuperscript{nd} Given Name Surname}
%\IEEEauthorblockA{\textit{dept. name of organization (of Aff.)} \\
%\textit{name of organization (of Aff.)}\\
%City, Country \\
%email address or ORCID}
%\and
%\IEEEauthorblockN{3\textsuperscript{rd} Given Name Surname}
%\IEEEauthorblockA{\textit{dept. name of organization (of Aff.)} \\
%\textit{name of organization (of Aff.)}\\
%City, Country \\
%email address or ORCID}
%\and
%\IEEEauthorblockN{4\textsuperscript{th} Given Name Surname}
%\IEEEauthorblockA{\textit{dept. name of organization (of Aff.)} \\
%\textit{name of organization (of Aff.)}\\
%City, Country \\
%email address or ORCID}
%\and
%\IEEEauthorblockN{5\textsuperscript{th} Given Name Surname}
%\IEEEauthorblockA{\textit{dept. name of organization (of Aff.)} \\
%\textit{name of organization (of Aff.)}\\
%City, Country \\
%email address or ORCID}
%}

\maketitle

\begin{abstract}
Transformers have transformed the field of natural language processing. This performance is largely attributed to the use of stacked ``self-attention" layers, each of which consists of matrix multiplies as well as softmax operations. As a result, unlike other neural networks, the softmax operation accounts for a significant fraction of the total run-time of Transformers. To address this, we propose \name{}, a hardware-friendly softmax design. \name{} consists of base replacement, low-precision softmax computations, and an online normalization calculation.
We show \name{} results in 2.35x the energy efficiency at 0.90x the size of a comparable baseline, with negligible impact on network accuracy.
\end{abstract}

\begin{IEEEkeywords}
neural network accelerators, hardware/software codesign, Transformers
\end{IEEEkeywords}

\putsec{Introduction}{introduction}
Transformer neural networks have recently become an extremely important deep learning (DL) workload, achieving state-of-the-art performance in a number of natural language processing tasks (NLP) ~\cite{attention}. These networks are characterized by their use of Transformer layers, which utilize self-attention, an attention mechanism that relates different symbols within a sequence in order to compute a representation of the sequence. 
%Self-attention consists of a set of fully-connected layers to generate query ($Q$), key ($K$), and value ($V$) activation matrices. The query and key are then multiplied together, and a softmax is applied to generate a normalized weighting. Finally, this weighting is multiplied by the value matrix and the result of the self-attention block is passed to the next layer.

Based on the success of the self-attention mechanism, the Transformer network and its later variants have quickly come to dominate the field of natural language processing. This success is not just limited to NLP tasks; Transformer-based networks have recently begun to also show tremendous promise on tasks previously dominated by Convolutional Neural Networks (CNNs), such as image recognition \cite{vt}. 
%Indeed, Transformer-based networks’ success in the visual domain prompted well-known ML research Oriol Vinyals to remark “Farewell convolutions”.

\begin{figure}[h]
    \centering
	\includegraphics[width=0.8\columnwidth]{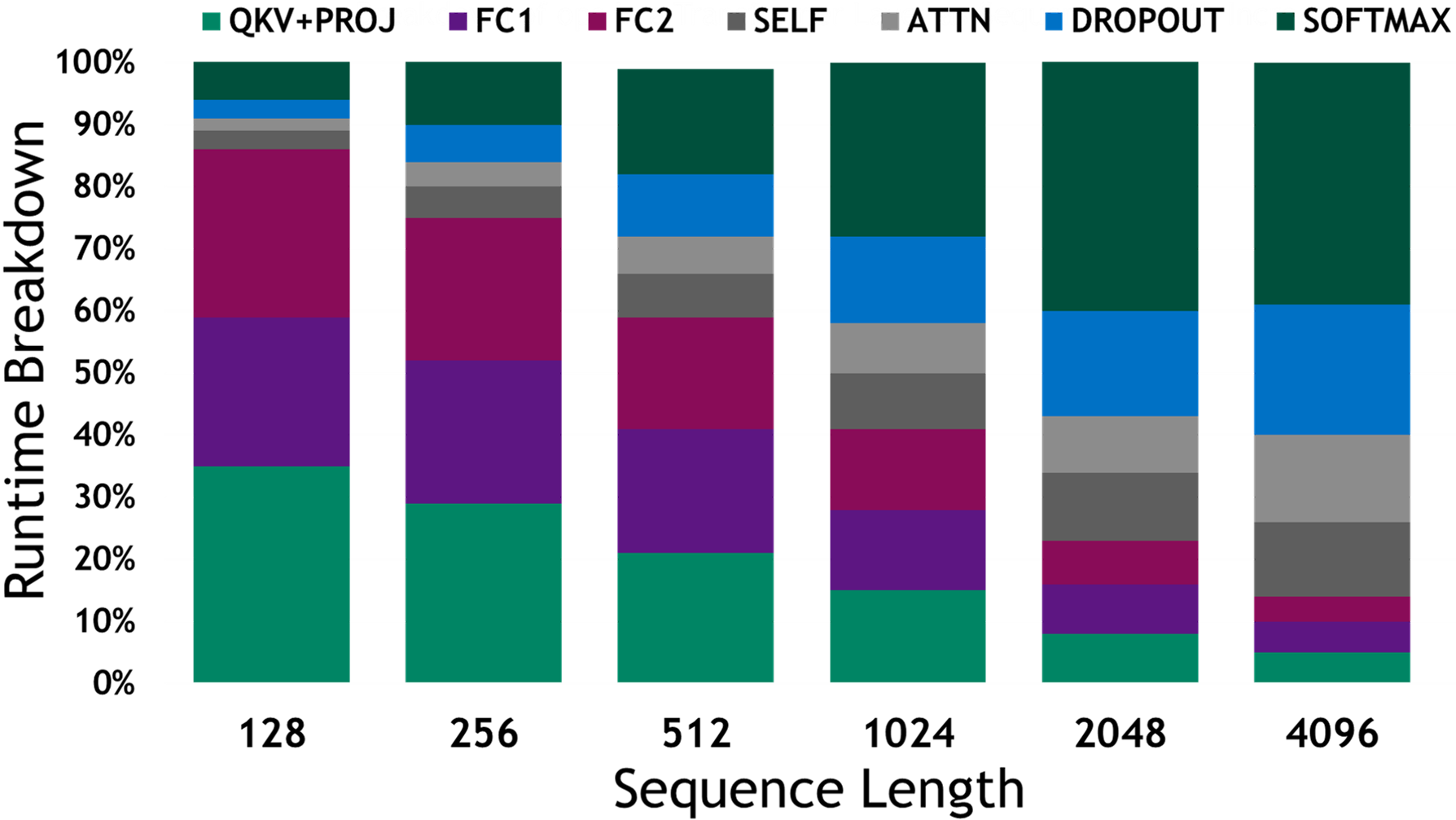}
	\caption[Softmax contributes more to runtime as sequence length increase]{Runtime breakdown for BERT-Large on a Volta GPU. Softmax contributes a larger fraction of the run time in Transformers than other DNNs, particularly at the longer sequence lengths seen in state-of-the-art networks}
	\label{fig:motivation}
	\vspace{-0.1in}
\end{figure}

However, the functional performance of Transformers comes at a cost. These networks are quite large, spanning hundreds of millions to hundreds of billions of parameters in recent networks such as BERT \cite{bert}, Megatron \cite{megatron}, GPT-2 \cite{gpt2} and GPT-3 \cite{gpt3}. These networks are continuing to grow in size; OpenAI's GPT-3 \cite{gpt3} has 175B parameters and an input sequence length of 2048 tokens in comparison to GPT-2's 1.5B parameters and sequence length of 1024. Looking beyond the high memory and compute overheads associated with large models, Transformer networks also have a unique mix of computations, as each attention layer of a Transformer network consists of softmax and dropout operations in addition to the standard matrix multiply-based fully-connected layers. As shown in \figref{fig:motivation}, these attention operations, particularly the softmax computation, represent a large fraction of  runtime, especially at the longer sequence lengths found in more recent state-of-the-art networks. 
%The growing dominance of Transformer-based networks has important implications for future hardware requirements.

%----------------------------------------------------------------------------------
%-------------------------------------------------------------------------------

%To date, however, optimizing the softmax function has been neglected in favor of focusing on more efficient linear algebra operations, due to the dominance of matrix multiply in workloads such as CNNs, while softmax contributes a negligible amount to the runtime of those workloads.

Previous DL inference accelerators have focused on CNNs, MLPs and LSTMs, which are dominated by matrix-multiply operations. Relatively little attention has been given to the acceleration of softmax, since it contributes a negligible amount to the runtime of these networks. With the growing importance of Transformers ({\em e.g.}, inference for conversational AI), it has become important to improve the performance of the softmax operation, which is the focus of this paper. To that end, we make the following contributions:
\begin{itemize}
\item We propose \name{}, a hardware-friendly softmax algorithm that consists of base replacement, low-precision softmax computations, and an online normalization calculation
\item Taking advantage of the fine-tuning paradigm of Transformer-based networks, we utilize \name{}-aware finetuning to reduce the accuracy loss incurred by our hardware-friendly algorithm while introducing no additional training overhead
\item We detail the microarchitecture necessary to implement \name{} in an inference accelerator
\item We demonstrate that \name{} achieves a 2.35x more energy efficient implementation while using 0.90x the area in a 7nm FinFET technology, with negligible impact on network accuracy
\end{itemize}

\begin{figure}[h]
	\centering 
	\includegraphics[width=\columnwidth]{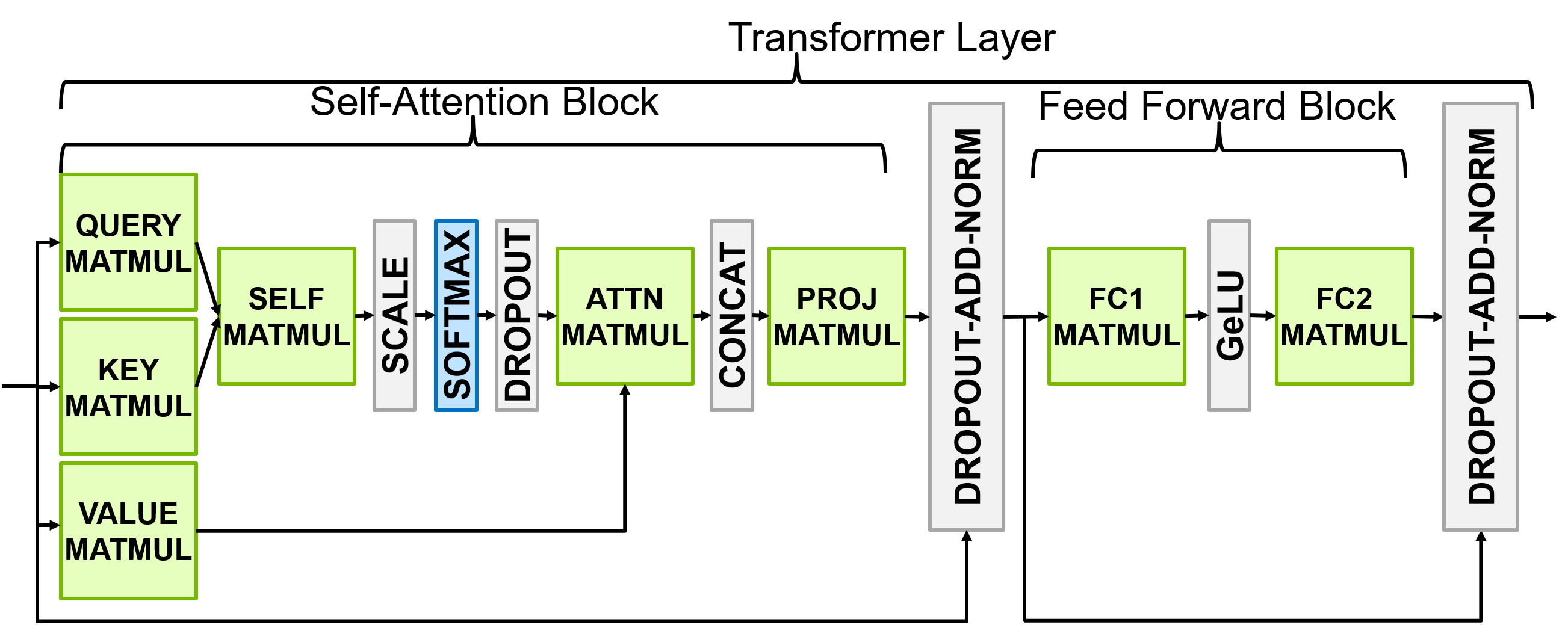}
	\caption[Overview of a Transformer Layer]{A Transformer layer consists of a multi-headed self attention block and a feedforward block. Of particular note is the use of softmax as a crucial operation in the self-attention block.}
	\label{fig:transformerlayer}
	\vspace{-0.1in}
\end{figure}

\putsec{Preliminaries and Related Work}{background}
In this section, we first describe the computations and bottlenecks seen in Transformer-based networks, followed by a brief overview of recent efforts to address these bottlenecks.
\putsubsec{Transformers}{transformers}
A Transformer network is a deep neural network (DNN)
that consists of one or more embedding layers, followed by multiple Transformer layers, and finally a task-specific final layer that is added when fine-tuning for the given task. The Transformer layer is the main algorithmic innovation in Transformer networks; for brevity, we will mainly focus on this layer, although our experimental evaluations are performed on complete networks. Transformer layers consist of a multi-headed attention block followed by a feed-forward block, as shown in \figref{fig:transformerlayer}.

%----------------------------------------------------------------------------------
%-------------------------------------------------------------------------------

The attention block applies three linear transformations to the input vector in order to obtain three new matrices: the query matrix ($Q$), the key matrix ($K$), and the value matrix ($V$). The query matrix and key matrix are then multiplied together and the result is scaled to the number of feature dimensions. Softmax is then applied to the resultant matrix, followed by dropout, resulting in the self-attention matrix $A$. This matrix is finally multiplied by the value matrix $V$. These operations can be repeated multiple times in parallel, resulting in multi-headed attention. The results of each head are concatenated together before being passed through multiple fully-connected layers in the feed forward stage. After both the multi-headed attention stage and feed forward stage, there is a set of dropout-add-norm layers.

%We make two main observations. First, each Transformer layer consists of six different sets of learnable parameters-- and each network consists of a multitude of these layers. This results in an extremely large number of parameters, contributing to large memory and computation requirements. Second, in addition to these matrix multiplications, Transformer layers also contain less common operations such as softmax and dropout. As shown in \figref{fig:motivation}, these non-matrix multiply operations begin to become larger bottlenecks, especially as recent Transformer-based networks have moved towards increasing sequence lengths, with the GPT family of networks moving from a sequence length of 512 to 2048 in the most recent version \cite{gpt3}.
We note that, in addition to the typical matrix multiplications, \emph{each} Transformer layer also contains less common operations such as softmax and dropout. As shown in \figref{fig:motivation}, these non-matrix multiply operations begin to become larger bottlenecks, especially as recent Transformer-based networks have moved towards increasing sequence lengths, with the GPT family of networks moving from a sequence length of 512 to 2048 in the most recent version \cite{gpt3}.

\putsubsec{Softmax Bottlenecks}{bottlenecks}
Existing hardware architectures for DL, which have been designed based on the characteristics of DNNs such as CNNs, MLPs and LSTMs, have largely neglected the softmax operation. This is because softmax is typically used only as the last layer in these networks in order to generate the final probabilities used in classification tasks, and thus represents only a small fraction of computation time and energy. However, this is no longer true for Transformer networks, which use softmax as a key component of the attention mechanism. For these networks, softmax can become a significant bottleneck, as shown in \figref{fig:motivation}.
The softmax operation is inefficient in current hardware for two main reasons. First, softmax requires the use of the exponential function. Exponential functions tend to require large look-up table (LUTs) to compute the result through the use of Taylor expansions. This is particularly true for general-purpose hardware such as CPUs and GPGPUs, which cater to exponential computations with high accuracy requirements due to their use in various scientific computing applications. This large area and power overhead makes it difficult to instantiate a large number of these units. Second, in order to improve training stability, deep neural networks typically use a numerically stable softmax, which subtracts the max of the vector on which softmax is being performed in order to ensure that the result does not blow up to infinity. However, this stability comes at a cost, as calculating the max introduces an additional pass through the vector, incurring latency and memory overheads.
\putsubsec{Related Work}{related}
%In this subsection, we first discuss efforts to address the large size of Transformer-based networks. Next, we discuss efforts to specifically address the softmax operation.
%\putsec{Related Work}{related}
%Intro sentence.

\noindent\textbf{Accelerating Transformers.} Most recent works have tried to make Transformers more efficient by targeting their large number of parameters, rather than targeting the softmax operation. This is commonly performed through the creation of smaller networks through techniques such as knowledge distillation \cite{distilbert}, inductive biases \cite{StarTransformer}, or approximations \cite{axtrans}. These optimizations are orthogonal to our effort, as \name{} can still be used in these smaller, more efficient networks. There have also been a few efforts towards more efficient Transformers through the use of lower-precision computation \cite{q8bert, distilbert}, with the most recent efforts also targeting quantization of the softmax operation \cite{fullyqt, integertransformer}. However, as these works are software-only, there are no actual gains in performance from their softmax quantization techniques. This is because full-precision special function units are still used for the exponential and division calculations-- in fact, there may actually be a slight performance penalty due to overheads in casting between data types. Our work, in contrast, is able to exploit low-precision in both software and hardware, unlocking actual performance gains. 

\noindent\textbf{Accelerating Softmax.} There have also been a few recent efforts that target the softmax operation directly \cite{ax-softmax-layer, efficient-softmax-hw, a3, precision-adjustable-softmax}. These efforts propose various low-precision exponential and division units, such as a variable precision softmax unit generator \cite{precision-adjustable-softmax}, a group lookup table-based exponential unit \cite{efficient-softmax-hw}, a split high-bits/low-bits exponential unit \cite{a3}, and approximate units \cite{ax-softmax-layer}. \name{} differs from these previous efforts for two main reasons. First of all, unlike \name{}, none of these efforts address the additional explicit pass required for computing the maximum, necessary for the numerically-stable softmax used in deep learning. Further, these efforts do not utilize base substitution, {\em i.e.}, their low-precision softmax implementations target the calculation of the natural exponential, rather than using base two as in our work.

\begin{figure}[htb]
    \centering
    \includegraphics[width=\columnwidth]{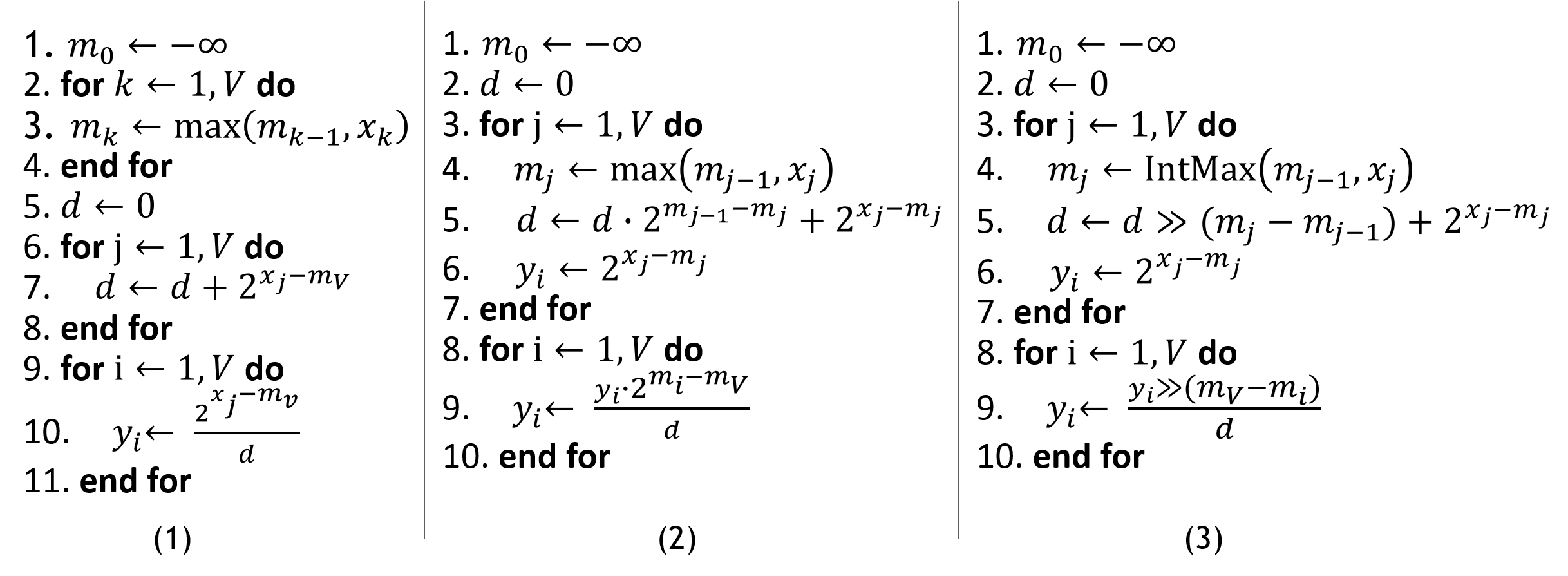}
	\caption[Overview of Proposed Hardfriendly-algorithm]{The algorithmic changes proposed in \name{} consist of: (1) replacing $e^x$ with a low-precision implementation of $2^x$, (2) replacing an explicit pass to calculate the max with an online version, and (3) replacing the maximum function with an integer-based version to simplify the renormalization calculations.  }
    \vspace{-0.1in} 
	\label{fig:algo}
\end{figure}

In summary, our work goes well beyond previous efforts by comprehensively optimizing softmax computations through base replacement, reduced precision and online normalization.
\putsec{\name{} Software}{software}
The \name{} algorithm, described in \figref{fig:algo}, is comprised of four enhancements over the standard softmax. First, we propose switching the base used in the exponential calculation from Euler's number, $e$, to $2$. Second, we replace full precision computations with fixed-point, low precision calculations-- including for exponentiation and division. Next, we use a hardware-friendly online normalization scheme to avoid an additional explicit pass to calculate the max.
%Finally, we note that the first two aspects of \name{} may introduce errors into the computation, impacting network accuracy. To mitigate this error, we propose \name{}-aware finetuning, wherein a non-\name{} based pretrained model is fine-tuned for a specific task, using \name{} in place of the standard softmax layer. We note that this fine-tuning for downstream tasks is already required for Transformer-based networks, so we are not introducing additional overheads; we are simply making the required fine-tuning \name{}-aware. In summary, a Transformer-based model is first \emph{pre-trained} using the standard, high-precision softmax, \emph{fine-tuned} for a specific downstream task in a \name{}-aware manner to mitigate accuracy loss, and deployed for \emph{inference} using \name{}.
Finally, we propose \name{}-aware finetuning, wherein a non-\name{} based pretrained model is fine-tuned for a specific task, using \name{} in place of the standard softmax layer, in order to account for errors introduced by our first two techniques. We note that this fine-tuning for downstream tasks is already required for Transformer-based networks, so we are not introducing additional overheads; we are simply making the required fine-tuning \name{}-aware. In summary, a Transformer-based model is first \emph{pre-trained} using the standard, high-precision softmax, \emph{fine-tuned} for a specific downstream task in a \name{}-aware manner to mitigate accuracy loss, and deployed for \emph{inference} using \name{}.

% Moving hardware figure here for better figure placement
\begin{figure*}[htb]
	\centering 
	\includegraphics[width=\textwidth]{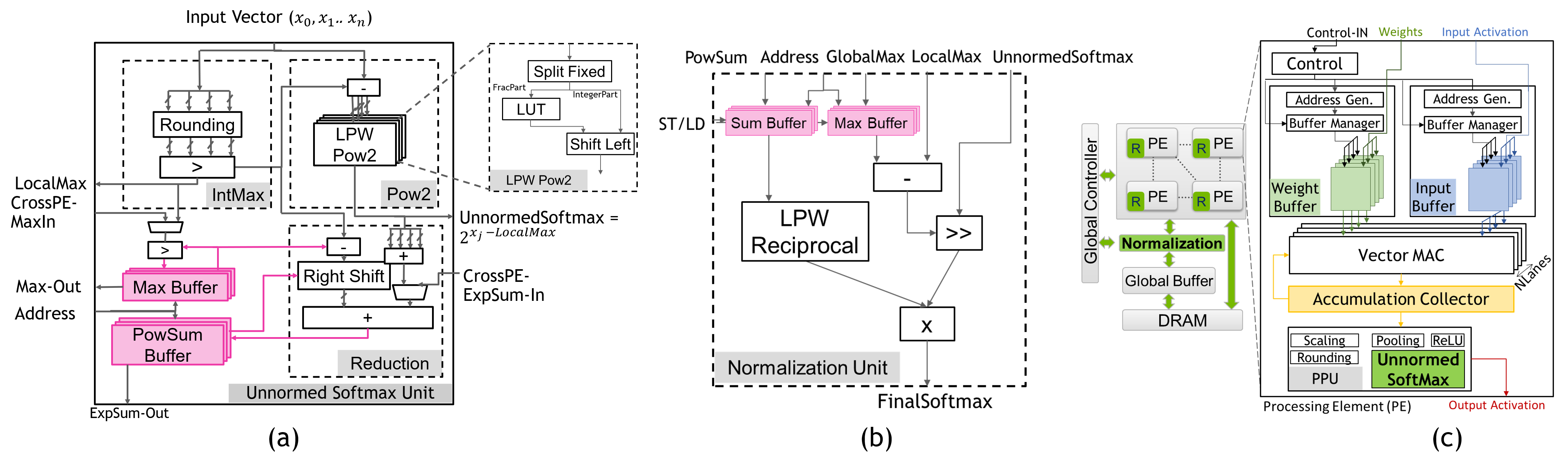}
	\caption[Diagram for All Units]{(a) The Unnormed Softmax Unit determines the local max, performs the power of 2 calculation using the current max, and accumulates the denominator. (b) The Normalization Unit performs the renormalization of the numerator, as well as the final division of the numerator by the accumulated sum. (c) In an example accelerator \cite{magnet}, the Unnormed Softmax can be integrated into the post-processing vector unit on a per PE basis, while the Normalization Unit can be shared across multiple PEs and integrated between the PEs and the Global Buffer}
	\label{fig:units}
	\vspace{-0.1in}
\end{figure*}
%\begin{figure*}[htb]
%    \centering
%    \includegraphics[width=0.75\textwidth]{softermax_algo}
%	\caption[Overview of Proposed Hardfriendly-algorithm]{The algorithmic changes proposed in \name{} consist of: (1) replacing $e^x$ with a low-precision implementation of $2^x$, (2) replacing an explicit pass to calculate the max with an online version, and (3) replacing the maximum function with an integer-based version to simplify the renormalization calculations.  }
%    \vspace{-0.1in} 
%	\label{fig:algo}
%\end{figure*}

\putsubsec{Base Replacement}{base}
Softmax is used in DNNs for three reasons. First, it produces a probability distribution which is useful for classification as well as in self-attention, where a probability can be thought of as a weighting. Second, it is differentiable and therefore can be used with gradient descent-based optimization. Finally, the use of the natural exponential provides a non-linear function wherein small differences in input values are exaggerated, leading to much higher probabilities for higher scores. 
We note that the use of the natural exponential results in overheads for both specialized and general-purpose hardware. In specialized hardware, it is usually cheaper to implement the natural exponential using a power of two compute unit coupled with a dedicated hardware multiplier for performing the base $e$ to base two conversion. Similarly, in general-purpose hardware, there is software overhead to perform this conversion.
We further note that a base two exponential still satisfies the three desirable properties of softmax listed above, while allowing for a more hardware-friendly implementation.
Therefore, \name{} uses a base-two based softmax. We demonstrate that this substitution leads to negligible loss in accuracy when \name{}-aware fine-tuning is performed. 
\putsubsec{Low-precision Softmax Operations}{lowprec}
Even with the above base replacement, softmax now consists of a power of two calculation, an accumulation, and a division. 
With 32-bit single-precision floating point, these units can be expensive in terms of area and power. Part of the reason these units are so expensive is the need for highly accurate results, particularly in general purpose computing platforms that cater to a wide range of applications. DNNs, by contrast, are quite resilient to errors introduced through low precision computing.
With this in mind, we propose performing all of the softmax compute operations - exponential, accumulation, and division - in low precision. We note that this is only possible through custom hardware, as commodity hardware platforms do not support low precision special function units like power and division.
Again, Softermax-aware fine-tuning minimizes accuracy loss as a result of these low-precision operations.

\putsubsec{Hardware-friendly Online Normalization}{onlinenorm}
In the DNN context, softmax is typically used in a “numerically stable” version, wherein the input vector is preprocessed by subtracting the maximum of the vector from every element. This helps with training stability, but at the cost of introducing an additional pass through the input vector.
Prior work \cite{onlinenorm} has proposed an “online normalizer” calculation, where the max is calculated continuously along with the normalization value ({\em i.e}, the summation for the denominator). In this case, the maximum of the vector is not subtracted before applying the exponential; rather, it is the maximum of the vector \textit{up to this point}. Thus, the current running sum must be renormalized when a new max is found. To see why this is the case in a concrete example, assume we are processing the vector [2,1,3]. For the first element, the running max is two and thus the running sum will be $d=2^{2-2}=1$. For the second element, the running max is still two and thus the running sum will be $d=1+2^{1-2}=1.5$. For the final element, however, we encounter a new maximum value of three. We cannot simply add to the existing sum as is (\emph{i.e.}, $d\neq1.5 + 2 ^{3-3} = 2.5$) since the previous accumulations were computed with a different point of reference for the numerically stable exponentiation. We must instead renormalize the running sum to account for this by multiplying by $2^{OldMax-NewMax}$, so that $d=1.5\times2^{2-3}+2^{3-3}=0.75+1=1.75$. Note that this result is the same as if we had computed the accumulation using the true global maximum from the start: $d=2^{2-3}+2^{1-3}+2^{3-3}=1.75$. For a mathematically rigorous proof of this algorithm, we refer the reader to \cite{onlinenorm}.

We propose a simple co-design modification in order to make the online calculation more hardware-friendly. Specifically, we note that, unlike the original online normalization algorithm, we use a base two implementation. With this in mind, we switch the max function with an integer max function, ensuring that the difference between maxes is always an integer. Since the renormalization operation is multiplying by $2^{OldMax-NewMax}$ and $OldMax-NewMax$ is guaranteed to be an integer, the renormalization hardware can therefore be realized simply using a shifter.

\putsec{\name{} Hardware}{hardware}
%-------------------------------------------------------------------------------
%\begin{figure*}[htb]
%	\centering 
	%\includegraphics[width=\textwidth]{units_combined_horizontal}
	%\caption[Diagram for All Units]{(a) The Unnormed Softmax Unit determines the local max, performs the power of 2 calculation using the current max, and accumulates the denominator. (b) The Normalization Unit performs the renormalization of the numerator, as well as the final division of the numerator by the accumulated sum. (c) In an example accelerator \cite{magnet}, the Unnormed Softmax can be integrated into the post-processing vector unit on a per PE basis, while the Normalization Unit can be shared across multiple PEs and integrated between the PEs and the Global Buffer}
	%\label{fig:units}
%	\vspace{-0.1in}
%\end{figure*}
%%-------------------------------------------------------------------------------

%In this section, we describe in detail the implementation of \name{} in hardware.
We propose two compute units, {\em viz.} the Unnormed Softmax unit and the Normalization unit, to realize \name{} (\figref{fig:units}). The Unnormed Softmax unit calculates the local maximum, performs the exponential, and accumulates the denominator (lines 4-6 in the final algorithm in \figref{fig:algo}). The Normalization Unit performs the final renormalization of the numerator and the division (lines 9-10). We provide greater detail of the implementation of these units in the subsections below. We also detail how these units may be integrated in an existing DNN inference accelerator.

\putsubsec{Unnormed Softmax Unit}{unnormed}
%The Unnormed Softmax Unit is designed to be integrated into an architecture which outputs multiple elements of a single output row at one time, which are then accumulated.
The Unnormed Softmax unit, shown in \figref{fig:units}(a), consists of three subunits: the IntMax unit, the Power of Two unit, and the Reduction unit.

\noindent\textbf{IntMax Unit:} The IntMax Unit operates on a slice of an output vector. It applies a ceiling function to each element in parallel before finding the max of the vector slice, thereby implementing an IntMax instead of simply max. 

\noindent\textbf{Power of Two Unit:} Power of two is implemented by decomposing the fixed point input into integer and fractional parts. The fractional part is implemented as a linear piece-wise function (LPW) applied to the range [0, 1), defined below.

\begin{equation*}
\begin{gathered}
    \label{eq:pwlpow}
    x_{scaled} = frac(x << 2) //\ 4\ segments\ in\ LPW \\
    lpw = m_{lut}[int(x_{scaled})]*frac(x_{scaled}) + c_{lut}[int(x_{scaled})] %\\
    %pow2 = pwl \cdot 2^{int(x)}
\end{gathered}
\end{equation*}

In our implementation, we use four segments in the LPW equation, while modern general purpose hardware typically utilize between 64-128 entries, a considerable overhead. To account for the four segments, our implementation (described above) requires a shift left by two (multiplication by four) in the first line  The fractional part of this scaled value ($frac(x_{scaled})$) is then multiplied by the output of the $m$ LUT. We note that, in cases where the input has less than two decimal places ({\em i.e.}, the input has two or fewer fractional bits), the fractional part will always be zero. Hence, the $m$ LUT is unused and the LPW implementation of the fractional part is simply:

\begin{equation*}
    lpw = c_{lut}[int(x_{scaled})]
\end{equation*}

Either way, the output of the LPW is then shifted by the integer part to obtain the final result.

\noindent\textbf{Reduction Unit:} The Reduction unit receives the $UnnormedSoftmax$ from the power of two unit and reduces it using a summation tree. It also reads from buffers in case the output vector is larger than can be computed in one slice. In this case, the current largest max for the row is read from the buffer along with the current running sum for that row. The current largest max is compared to the $LocalMax$ found and the running sum is renormalized as needed using a shifter when there is a difference between the local max and the current determined max for a row, as required by the online normalization algorithm. The renormalized running sum and the local sum are then added together to obtain the new running sum for the row.

\putsubsec{Normalization Unit}{normunit}
The Normalization Unit performs the renormalization of the numerator as well as the division to obtain the final result. As a result of the integer max used in \name{}, it is guaranteed that the difference between the local maximum and the global maximum is an integer. Therefore, due to the base change used in \name{}, the renormalization of the numerator can be implemented simply using a shifter. The division is implemented using a linear piece-wise reciprocal unit, followed by an integer multiplier.

%----------------------------------------------------------------------------------
%\begin{figure}[htb]
%	\centering 
%	\includegraphics[width=0.75\columnwidth]{normalization_unit}
%	\caption[Diagram for Normalization Unit]{The Normalization unit performs the renormalization of the numerator, as well as the final division of the numerator by the accumulated sum.}
%	\label{fig:normalization_unit}
%	\vspace{-0.1in}
%\end{figure}
%-------------------------------------------------------------------------------

\putsubsec{Accelerator Integration}{integration}
The UnnormedSoftmax Unit is designed to integrate into tensor processing hardware, present in GPU tensor cores \cite{volta}, TPUs \cite{tpu}, or other dedicated DNN inference accelerators.  For example, using the MAGNet architecture \cite{magnet} as a baseline, the unit can be integrated into its post-processing unit (PPU), which performs operations such as pooling and ReLu. Ideally, the UnnormedSoftmax Unit should be sized such that it matches the MAC throughput, fully exploiting the low-overhead hardware enabled by Softermax.
The Normalization Unit can be introduced between the processing tile and global memory to complete the softmax off of the critical path.

%%----------------------------------------------------------------------------------
%\begin{figure}[htb]
%	\centering 
%	\includegraphics[width=\columnwidth]{arch_overview}
%	\label{fig:arch_overview}
%	\caption[Diagram for Accelerator-Level Integration]{An example of how the \name{} units can be integrated into an existing neural network accelerator \cite{magnet}. The Unnormed Softmax is integrated into the post-processing vector unit on a per PE basis. The Normalization unit is shared across multiple PEs and integrated between the PEs and the Global Buffer.}
%	\vspace{-0.1in}
%\end{figure}
%%-------------------------------------------------------------------------------
\putsec{Experimental Setup}{setup}
\noindent\textbf{Software setup:} To evaluate the impact of \name{} on accuracy, we modified the PyTorch Huggingface library \cite{huggingface}. Our Huggingface implementation uses a 99.999\% percentile calibrator to generate the scale factors which are then used to to perform quantization-aware finetuning with 8-bit weights and activations \cite{integerquant}. We further augment our implementation with custom forward/backward passes for each of the fixed point softmax operations: power of 2, reciprocal, etc. The forward passes faithfully implement the fixed point, low precision computations, while the backward passes use the straight through estimator (STE) for the conversions to/from fixed point. The bitwidths for each operation are summarized in Table~\ref{table:bitwidths}. We note that the input and output of our \name{} are 8-bits each, allowing for easy integration into existing 8-bit integer vector MAC datapaths, such as those found in modern GPU tensor cores and DNN inference accelerators.

\begin{table}[htb]
\renewcommand{\tabcolsep}{4pt}
\renewcommand{\arraystretch}{2.5}
\caption{Summary of \name{} Bitwidths,  Q(Int., Frac.)}
\fontsize{9}{7.2}\selectfont
\begin{tabular}{cccccc}
\hline
\hline
\textbf{Inp.}   & \textbf{LocalMax} & \textbf{Unnormed} & \textbf{PowSum}   & \textbf{Recip.}   & \textbf{Outp.}    \\ \hline \hline
Q(6,2)          & Q(6,2)            & Q(1,15)           & Q(10, 6)          & Q(1,7)            & Q(1, 7)           \\ \hline
\end{tabular}
\label{table:bitwidths}
\end{table}

\noindent\textbf{Hardware setup:} To quantify the area and power overheads associated with \name{}, we implemented the proposed hardware units using high-level synthesis and integrated the units into an existing accelerator, MAGNet \cite{magnet}. We compare our \name{} implementation to a standard 16-bit floating point precision softmax implementation that uses Synopsys DesignWare components\cite{designware}. We note that this represents an optimistic baseline already, as current state of the art accelerators \cite{volta} use a full 32-bit precision implementation. 
\begin{table}[htb]
\renewcommand{\tabcolsep}{4pt}
\renewcommand{\arraystretch}{1.25}
\caption{Experimental Setup}
\begin{tabular}{cc}
\hline
\hline
\multicolumn{2}{c}{{\bf Design Tools}}\\
\hline \hline
\textbf{HLS compiler:} & Mentor Graphics Catapult HLS \\
\hline
\textbf{Verilog simulator:} & Synopsys VCS \\
\hline
\textbf{Logic synthesis:} & Synopsys Design Compiler Graphical \\
\hline
\textbf{Power analysis:} & Synopsys PT-PX \\
\hline \hline
\multicolumn{2}{c}{{\bf Design Parameters}}\\
\hline \hline
\textbf{Weight/Activation precision:} & 8 bits \\
\hline
\textbf{Accumulation precision:} & 24 bits \\
\hline
\textbf{VectorSize:} & 16~\textbar~32 \\
\hline
\textbf{NLanes:} & 16~\textbar~32 \\
\hline
\textbf{Input Buffer Size:} & 16KB~\textbar~32KB  \\
\hline
\textbf{Weight Buffer Size:} & 32KB~\textbar~128KB  \\
\hline
\textbf{Accumulation Collector Size:} & 6KB~\textbar~12KB  \\
\hline
\textbf{Technology Node:} & TSMC 7nm \\
\hline
\textbf{Supply Voltage:} & 0.67V \\
\hline
\hline
\end{tabular}
\label{table:magnet_params}
\end{table}

\putsec{Evaluation}{evaluation}
We split our evaluation of \name{} into two parts. First, we demonstrate that our proposals for making softmax more hardware-friendly do not have a negative impact on accuracy. Second, we demonstrate that these proposals do in fact result in more efficient hardware implementations.

\putsubsec{Impact on Accuracy}{accuracy}
%For our accuracy experiments, we compare against an eight-bit quantized baseline, which uses a 99.999\% percentile calibrator to generate the scale factors, which are then applied to the weights and activations. In the baseline, we utilize quantization-aware fine-tuning to fine tune an unsupervised BERT model for each task. For our \name{} results, we keep the eight-bit quantized weights and activations, and add in our \name{} changes that impact accuracy: low precision hardware and base two substitution. We use the bitwidths listed in Table~\ref{table:bitwidths}. We again utilize fine-tuning for each task. We note that hyperparameters are kept constant in both the baseline and \name{} training regimes. 
We use the setup described in the previous section to analyze the impact of our proposed \name{} implementation.

%\begin{figure*}[htb]
%    \centering
%    \includegraphics[width=0.9\textwidth]{accuracy_results}
%	\caption[Accuracy Impact of Softermax]{Softermax results in negligible loss in accuracy compared to an 8-bit quantized baseline which uses full-precision, numerically stable softmax.}
%    \vspace{-0.1in} 
%	\label{fig:accuracy}
%\end{figure*}
\begin{table*}[]
\centering
\renewcommand{\arraystretch}{1.25}
\caption{Accuracy results for \name{} and an eight-bit quantized baseline across SQuAD and GLUE tasks. \name{} incurs no average accuracy loss.}
\begin{tabular}{ccccccccccc}
\hline
\hline
                                                                               & \textbf{}          & \textbf{SQuAD} & \textbf{RTE} & \textbf{CoLA} & \textbf{MRPC} & \textbf{QNLI} & \textbf{QQP} & \textbf{SST-2} & \textbf{STS-B} & \textbf{MNLI} \\ \hline\hline
\multirow{2}{*}{\textbf{\begin{tabular}[c]{@{}c@{}}BERT\\ Base\end{tabular}}}  & \textbf{Baseline}  & 86.28          & 62.45        & 53.65         & 84.31         & 90.77         & 90.71        & 92.09          & 87.86          & 83.27         \\
                                                                               & \textbf{Softermax} & 85.86          & 64.26        & 56.76         & 84.07         & 90.41         & 90.83        & 92.20          & 87.78          & 83.80         \\ \hline
\multirow{2}{*}{\textbf{\begin{tabular}[c]{@{}c@{}}BERT\\ Large\end{tabular}}} & \textbf{Baseline}  & 89.40          & 65.70        & 59.58         & 86.03         & 92.09         & 91.24        & 92.89          & 89.39          & 85.87         \\
                                                                               & \textbf{Softermax} & 89.46          & 69.68        & 60.10         & 86.27         & 91.76         & 90.90        & 92.66          & 89.55          & 85.74         \\ \hline
\end{tabular}
\label{table:accuracy}
\end{table*}

Under these conditions, we find that \name{} results in negligible loss in accuracy. Table~\ref{table:accuracy} details the accuracy results on the BERT-Base and BERT-Large networks, across the SQuAD and GLUE tasks. To summarize, \name{} results in negligible impact on accuracy; the worst drop in accuracy is under 0.5\%, while the average accuracy actually \textit{increases} 0.9\% and 0.7\% across all tasks for BERT-Base and BERT-Large, respectively. This validates the use of \name{} as an acceptable replacement for full-precision, numerically-stable softmax in Transformer-based networks. 

\putsubsec{Impact on Hardware Efficiency}{hardware-eval}

\noindent\textbf{Compute Unit Level Analysis.} We first compare our proposed \name{} implementation to a DesignWare-based baseline \cite{designware} in isolation-- \emph{i.e.}, not integrated into the PE of a DNN accelerator. Specifically, we compare the Unnormed Softmax unit, as implemented using the techniques outlined in \name{}, to a unit in which the requisite units (max, exponential, accumulation) are implemented using DesignWare components. We consider a softmax workload with a sequence length of 384, as used in SQuAD dataset. We see that \name{} offers a much more efficient implementation, resulting in an Unnormed Softmax unit that is \texttt{4x} smaller and \texttt{9.53x} more energy efficient as demonstrated in Table~\ref{table:unit_level}. Similarly, the Normalization Unit is much more efficient than the baseline, resulting in a unit that is  \texttt{1.54x} smaller and \texttt{2.53x} more energy efficient.

\begin{table}[]
\centering
\caption{Softermax comparison to DesignWare-based softmax baseline for SQUAD dataset}
\renewcommand{\tabcolsep}{4pt}
\renewcommand{\arraystretch}{1.25}
\begin{tabular}{ccc}
\hline
\hline
\textbf{}                       & \textbf{Area ($um^2$)}    & \textbf{Energy (uJ)}  \\ \hline\hline
\textbf{Unnormed Softmax Unit}  & 0.25x                     & 0.10x                 \\ \hline
\textbf{Normalization Unit}     & 0.65x                     & 0.39x          \\ \hline
\textbf{Full PE}                & 0.90x                     & 0.43x                 \\ \hline
\end{tabular}
\label{table:unit_level}
\end{table}

\begin{figure}[h]
	\includegraphics[width=\columnwidth]{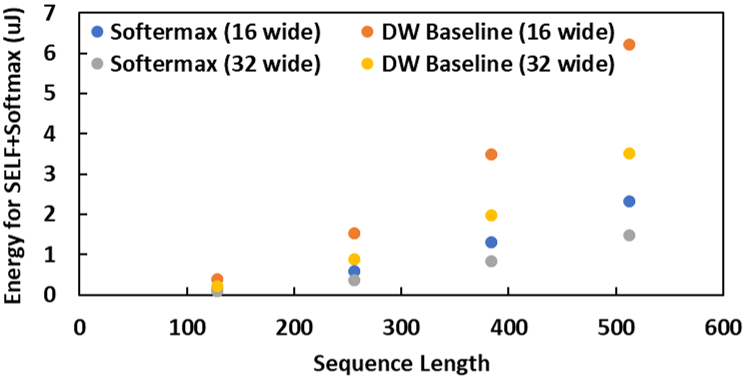}
	\caption[Softmax-based PEs are much more energy-efficient]{Energy consumption of a \name{}-based PE vs a DesignWare baseline for SELF+Softmax as the sequence length increases. We evaluate 16- and 32-wide implementations.}
	\label{fig:seq_sweep}
	\vspace{-0.1in}
\end{figure}
%-------------------------------------------------------------------------------

\noindent\textbf{PE-level Analysis.} Next, we compare \name{} to the DesignWare-based baseline when integrated into the PE of a DNN accelerator. Specifically, we integrated the Unnormed Softmax unit into a 32-wide MAGNet PE, with parameters as described in Table~\ref{table:magnet_params}. As shown in Table~\ref{table:unit_level}, we see that \name{} still offers a more efficient solution even when integrated into a PE and thus accounting for other, non-softmax related components such as the MAC units and various scratchpads; \name{} is \texttt{1.11x} more area efficient and \texttt{2.35x} more energy efficient than the baseline.

\noindent\textbf{Sequence Length Sweep.} Finally, we perform a sweep of the sequence length of the input vector, evaluating both 32-wide as well as 16-wide PE configurations. As shown in \figref{fig:seq_sweep}, \name{} scales much better than the baseline, both starting from a lower baseline and having a more shallow slope. This is crucially important in Transformer-based networks, which are trending towards longer and longer sequence lengths.
\putsec{Conclusion}{conclusion}
Transformer networks are an important emerging class of deep learning workloads, which differ computationally from other DNNs through their extensive use of softmax computations. Implementations of Transformers on current general-purpose and specialized hardware platforms are therefore limited by the time and energy
of softmax operations. To address this challenge, we proposed \name{}, a set of software and hardware optimizations to softmax operations in Transformer networks. Our implementation of \name{} indicates that it can lead to 4x area and 9.53x energy improvements over conventional softmax units, translating to 1.11x area improvement and 2.35x energy improvement for softmax computations within a state-of-the-art DNN accelerator, with no loss in accuracy compared to the quantized baseline.

%%%%%%%%% -- BIB STYLE AND FILE -- %%%%%%%%
\bibliographystyle{IEEEtran}
\bibliography{refs}
%%%%%%%%%%%%%%%%%%%%%%%%%%%%%%%%%%%%

\end{document}